\documentclass[referee]{aa} 
\usepackage{graphicx}
\begin{document}

\titlerunning{The LFs of the galaxy cluster EIS\,0048 at $z\sim 0.64$}

\title{UV--NIR Restframe Luminosity Functions of the
Galaxy Cluster EIS\,0048 at $z\sim 0.64$\thanks{Based on observations collected at the European Southern Observatory}}

\subtitle{}

\author{M. Massarotti \inst{1}
\and
G. Busarello \inst{1}
\and
F. La Barbera\inst{1}  
\and 
P. Merluzzi \inst{1}
}
\offprints{M. Massarotti}

\institute{I.N.A.F., Istituto Nazionale di Astrofisica 
Osservatorio Astronomico di Capodimonte, 
Via Moiariello 16, I-80131 Napoli, Italy\\
email: michele@na.astro.it
}

\date{Received ; accepted }

\abstract{We derive the galaxy luminosity functions in V-, R-, I-,
and K-bands of the cluster EIS\,0048 at $z \sim 0.64$ from data
taken at the ESO Very Large Telescope. 
The data span the restframe wavelength range from UV, which is sensitive to 
even low rates of star formation, to the NIR, which maps the bulk of the 
stellar mass.  

By comparing our data and previous results with pure luminosity 
evolution models, we conclude that bright ($M \leq M^*+1$) cluster galaxies
are already assembled at $z \sim 1$ and that star formation is almost
completed at $z \sim 1.5$.

}

   \maketitle
%

\section{Introduction}\label{INTRO}
One of the main issues in theoretical and observational research is to
understand the physical processes driving the formation and evolution
of bright massive galaxies in clusters, and to constrain the relative
time scales.

The evolution of the cluster galaxy sequence (the colour--magnitude
relation) has been studied for large cluster samples up to redshift $z
\sim 1$ (Aragon--Salamanca et al. \cite{AEC93}; Lubin \cite{LUB96};
Ellis et al. \cite{ESD97}; Stanford et al. \cite{SED98}, hereafter
SED98; Nelson et al. \cite{NGZ01}, hereafter NGZ01).  The results are
consistent with a monolithic collapse scenario (see e.g. Larson
\cite{L74}) in which galaxies form at high redshifts and subsequently
undergo passive evolution.  However, the evolution of colours only
inform on the epoch when the bulk of stellar mass formed, while
cluster bright galaxies ($M \leq M^*+1$, mostly early--type) could
have been also assembled recently ($z < 1$) from mergers of smaller
units, at least as long as the merging processes do not induce strong
star formation.

A different approach consists in the study of the evolution of the
cluster luminosity function (CLF). In particular, the near--infrared
(NIR) CLF can be used to assess the assembly history of galaxies,
because the NIR light mainly informs on galaxy mass.  From the
analysis of the values of $M^*$ in the K-band as a function of
redshift for a sample of 38 clusters in the range $0.1 < z < 1$, De
Propris et al. (\cite{DPSE99}, hereafter DPSE99) found a trend
consistent with passive luminosity evolution (PLE) (see also Nakata et
al. \cite{NAK01}, hereafter NKY01).  DPSE99 conclude that the mass
function of bright cluster galaxies is invariant at $z < 1$, and that
the assembly of those galaxies is largely complete by $z \sim 1$.
Spectroscopic data also show that massive galaxies already exist at
least up to $z=0.83$ (van Dokkum et al. \cite{vDF98}).

In this work we derive the CLFs for EIS\,0048 at $z \sim 0.64$ in the
V-, R-, I-, and K-band (UV to NIR restframe).  The cluster membership
has been assessed with the photometric redshift technique up to $M-M^*
\sim 1.5-2.5$ (according to the different depth of each band).  Since
no other selection has been applied, the samples are not biased toward
a particular galaxy population.

The paper is organized as follows. In Sect. 2 we introduce the
photometric data, discuss the background subtraction, the selection of
cluster members, the completeness of the samples, and obtain the CLFs.
In Sect. 3 we discuss the results in terms of galaxy formation and
evolution. In Sect. 4 we give a summary of the paper. In this work we
assume H$_0=70$ km s$^{-1}$ Mpc$^{-1}$, $\Omega_M=0.3$ and
$\Omega_{\Lambda} = 0.7$.

\section{Derivation of the luminosity functions}\label{DATA}

The photometric observations of the cluster of galaxies EIS\,0048 were
carried out at the ESO Very Large Telescope (VLT) during two observing
runs on August 2001. All the nights were photometric with excellent
seeing conditions.

The data include VRIK imaging taken with the FORS2 and ISAAC
instruments, respectively.  The VRI images consist of a single pointing
of $6.8 \! \times \! 6.8~\mathrm{arcmin}^2$ for each band, while for
the K-band a mosaic of four pointings covers a total area of $4.9 \!
\times \! 4.9~\mathrm{arcmin}^2$. Further details on the observations
and on the data reduction can be found in La Barbera et
al. (\cite{LBM03}, hereafter LBMI03).

\subsection{Magnitudes and completeness}\label{COMGAR}

Total magnitudes were computed by means of the software SExtractor
(Bertin \& Arnouts 1996). For each object we obtained Kron magnitudes
($\mathrm{m_K}$) within an aperture of diameter $\alpha \cdot
\mathrm{r_K}$, where $\mathrm{r_K}$ is the Kron radius (Kron 1980). We
chose $\alpha =2.2$, for which $\mathrm{m_K}$ is expected to enclose
92\% of the total flux, and we computed the total magnitudes by adding
0.08 mag to $\mathrm{m_K}$ (see LBMI03 for details).

\begin{figure}
\includegraphics[angle=0,width=15cm,height=15cm]{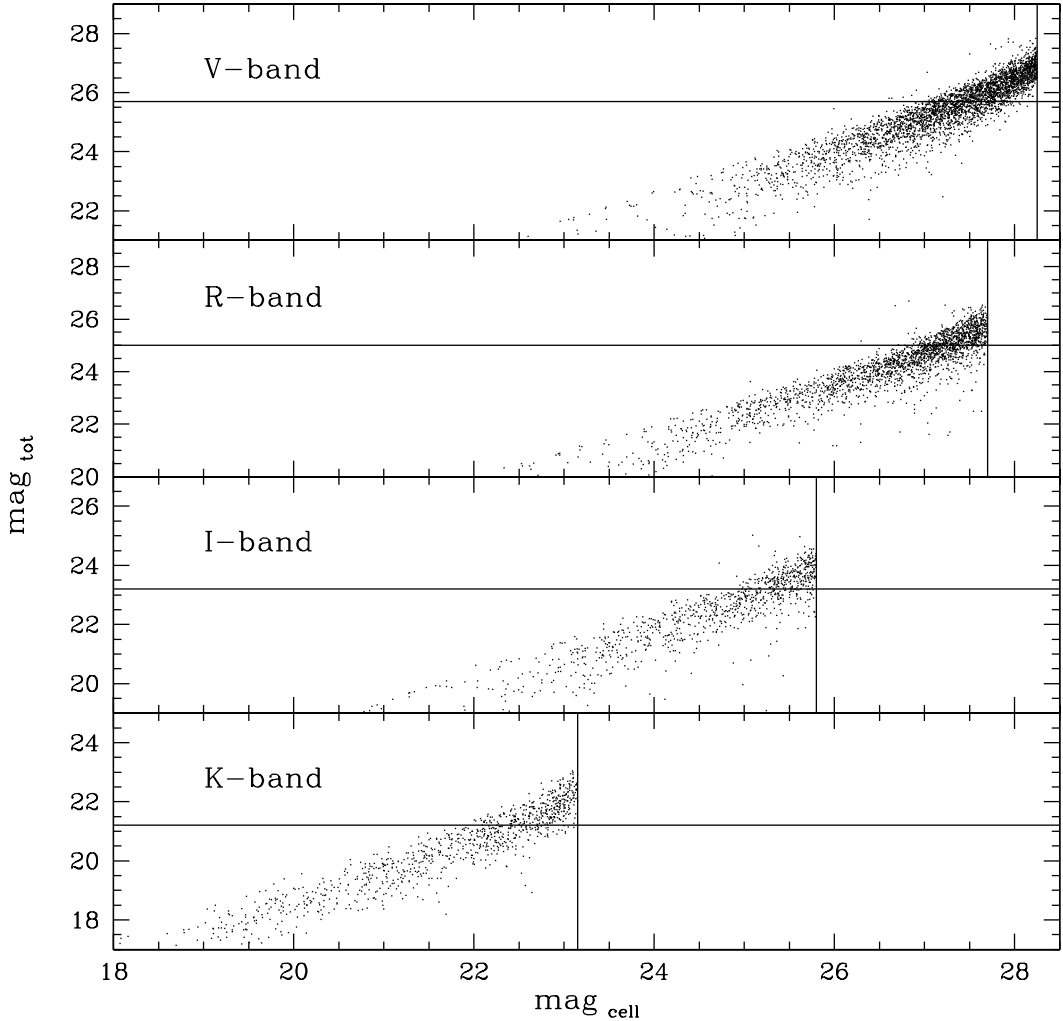}
\caption[]{Completeness limits of the photometric
catalogues. Total magnitudes (mag$_{tot}$) are plotted against the
magnitudes in the detection aperture (mag$_{cell}$). The vertical and
horizontal lines correspond to the detection threshold and to the
completeness magnitudes respectively.

\label{FIGCOMP}
}

\end{figure}

The formal completeness of the catalogues was estimated following the
method of Garilli et al. (\cite{GAR99}, see also Busarello et
al. 2002).  This method consists in the determination of the magnitude
at which the objects start to be lost since they are below the
brightness threshold in the detection cell 
\footnote{ In the present case, the detection cell is an aperture with
diameter of 5 pixels.}. To estimate the completeness limit, magnitudes
in the detection cell were compared to the total magnitudes. The
comparisons are shown in Fig.~\ref{FIGCOMP}, where the vertical lines
correspond to the detection threshold and the horizontal lines mark
the completeness limits (taking into account the scatter of the
relation). The completeness magnitudes are reported in
Table~\ref{TABCOMP}.

\begin{table}
\caption[]{The completeness magnitudes. $\mathrm{m^{c}}$ is the
completeness magnitude as estimated on the photometric catalogues
according to Garilli et al. (\cite{GAR99}). $\mathrm{m^{c}_{zp}}$
defines the central position of the faintest magnitude bin with
completeness higher than $50\%$ in the sample with photometric
redshifts (see Sect.\ref{BS}). The bin width is 0.5 mag.}
\label{TABCOMP}
$$
\begin{array}{cccc}
\hline
\noalign{\smallskip}
$$\mathrm{Band}$$ & $$\mathrm{m^{c}}$$ & $$\mathrm{m^{c}_{zp}}$$\\
& $$\mathrm{mag}$$ & $$\mathrm{mag}$$ & \\
\noalign{\smallskip}
\hline
\noalign{\smallskip}
$$\mathrm{V}$$ & 25.7 & 24.25 \\
$$\mathrm{R}$$ & 25.0 & 23.25 \\
$$\mathrm{I}$$ & 23.2 & 22.75 \\
$$\mathrm{K}$$ & 21.2 & 19.75 \\
\noalign{\smallskip}
\hline
\end{array}
$$
\end{table}

\subsection{Background subtraction}\label{BS}

The LFs of cluster galaxies at low and intermediate redshifts are
usually obtained by statistical subtraction of the background and
foreground contribution (as derived from a control field) to the
galaxy counts in the cluster field.  In this way all the photometric
data up to the magnitude limit can be used.  On the other hand, some
problems become severe as the cluster redshift increases:
a) cluster galaxies become fainter; b) the number of galaxies in
the control field becomes higher; c) clusters are younger and therefore
the number of cluster galaxies is smaller.  
The consequences are that the contrast of
cluster galaxies over the field decreases as a function of redshift,
and that the Poissonian fluctuations of the field as well as
field--to--field variations become higher than the signal to be
detected (see also NKY01).

To analyse this problem, we compare a known CLF translated to
different redshifts, with the uncertainties on field galaxy counts.
We use the LF of the rich cluster AC\,118 at $z=0.31$,
$N_{\mathrm{CLU}}(m)$, in the K-band from Andreon (\cite{AND01}), and
the galaxy counts $N_{\mathrm{HDFS}}(m)$ and $N_{\mathrm{CDF}}(m)$
from two different control fields: the Hubble Deep Field South (HDFS)
and the Chandra Deep Field (CDF) respectively, as obtained by Saracco
et al. (\cite{SGC01}).

We define the {\it cluster signal--to--noise ratio CSNR} as:
\begin{equation}
 CSNR(m)=\frac{N_{\mathrm{CLU}}(m)}{\sqrt{(N_{\mathrm{HDFS}}(m)-N_{\mathrm{CDF}}(m))^{2}+N_{\mathrm{CLU}}(m)}}
 \ ,
\label{EQCSNR}
\end{equation}
where $N_{\mathrm{CLU}}(m)$ is the signal we aim to measure and
$\sqrt{(N_{\mathrm{HDFS}}(m)-N_{\mathrm{CDF}}(m))^{2}+N_{\mathrm{CLU}}(m)}$
is an estimate of the uncertainties on the field galaxy counts plus the 
Poissonian errors on the CLF.

\begin{figure}
\includegraphics[angle=0,width=13cm,height=10cm]{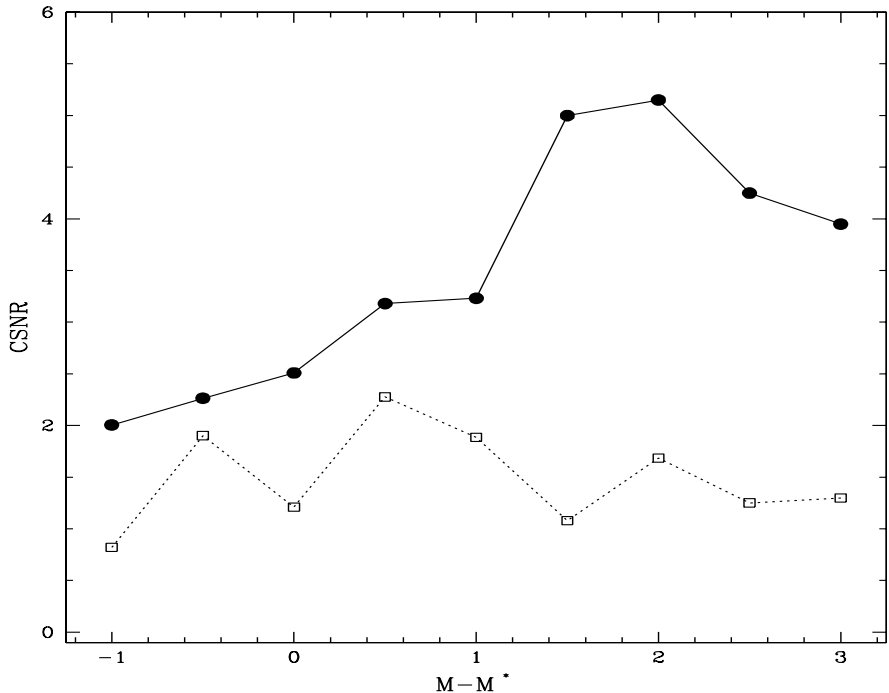}
\caption[]{ The cluster signal--to--noise ratio (see text and
Eq.~\ref{EQCSNR}) in a rich cluster at $z=0.31$ (filled dots and
continuous line) and at $z=0.64$ (open squares and dotted line).
\label{PSCSNR}
}
\end{figure}

We computed the $CSNR$ for AC\,118 at $z=0.31$ (filled circles and
continuous line in Fig.~\ref{PSCSNR}) and then shifted it to $z=0.64$
(open squares and dotted line) according to the PLE model described in
Sect.~\ref{DISC}.  As can be seen in Fig.~\ref{PSCSNR}, for $z=0.31$
$CSNR(m) > 2$ at all magnitudes and $CSNR(m) > 3$ at $m > M^{*}+0.5$
($M^{*}=15.3$, see Andreon \cite{AND01}).  At $z=0.64$, $CSNR(m) \le
2$ at all magnitudes, implying that the signal to be detected would be
almost equal to the noise.

An additional problem arises from two facts: a) AC\,118 is a cluster
richer than EIS\,0048; b) by shifting the LF of AC\,118 at a higher
redshift, we overestimated the actual number of detected galaxies.  As
consequence, our estimate of $CSNR(m)$ at $z=0.64$ must be considered
as an upper limit, and adopting the procedure of statistically
subtracting the background and foreground contribution from a control
field would prevent an accurate estimate of the CLFs.  It should also
be noted that we do not have access to a control field in the
immediate proximity of EIS\,0048 in order to minimize uncertainties
from field--to--field variations (as in NGZ01), and that we cannot
derive a cumulative LF by combining the signals from different
clusters at similar redshifts (as in DPSE99).

\begin{figure}
\includegraphics[angle=0,width=15cm,height=15cm]{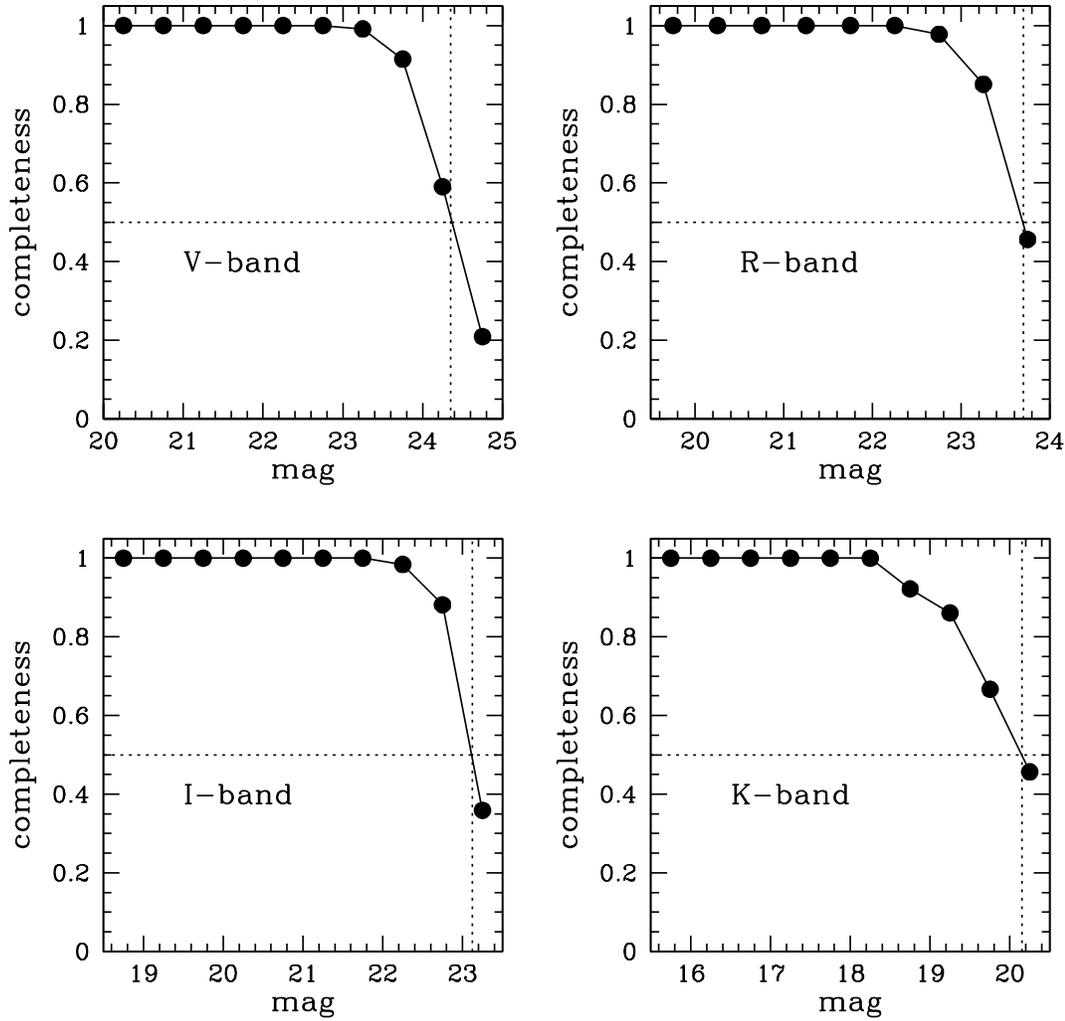}
\caption[]{The completeness as a function of the magnitude of the
samples used to derive the photometric redshift distribution of the
galaxies in the EIS\,0048 field. In each panel the vertical dotted
line indicates the magnitude corresponding to $50 \%$
completeness. The CLFs are computed up to the faintest magnitude bin
brighter than these limits.

\label{PSCOMP}
}
\end{figure}

Therefore, in order to increase the contrast of cluster galaxies over
field galaxies, the following approach was adopted. We first selected
the cluster members through the photometric redshift technique, as
detailed in LBMI03. In this way it is possible to isolate galaxies in
a narrow interval $\Delta z \pm 0.1$ around the cluster redshift, thus
greatly decreasing the noise produced by field galaxies, while leaving
the signal untouched.

\begin{figure}
\includegraphics[angle=0,width=15cm,height=15cm]{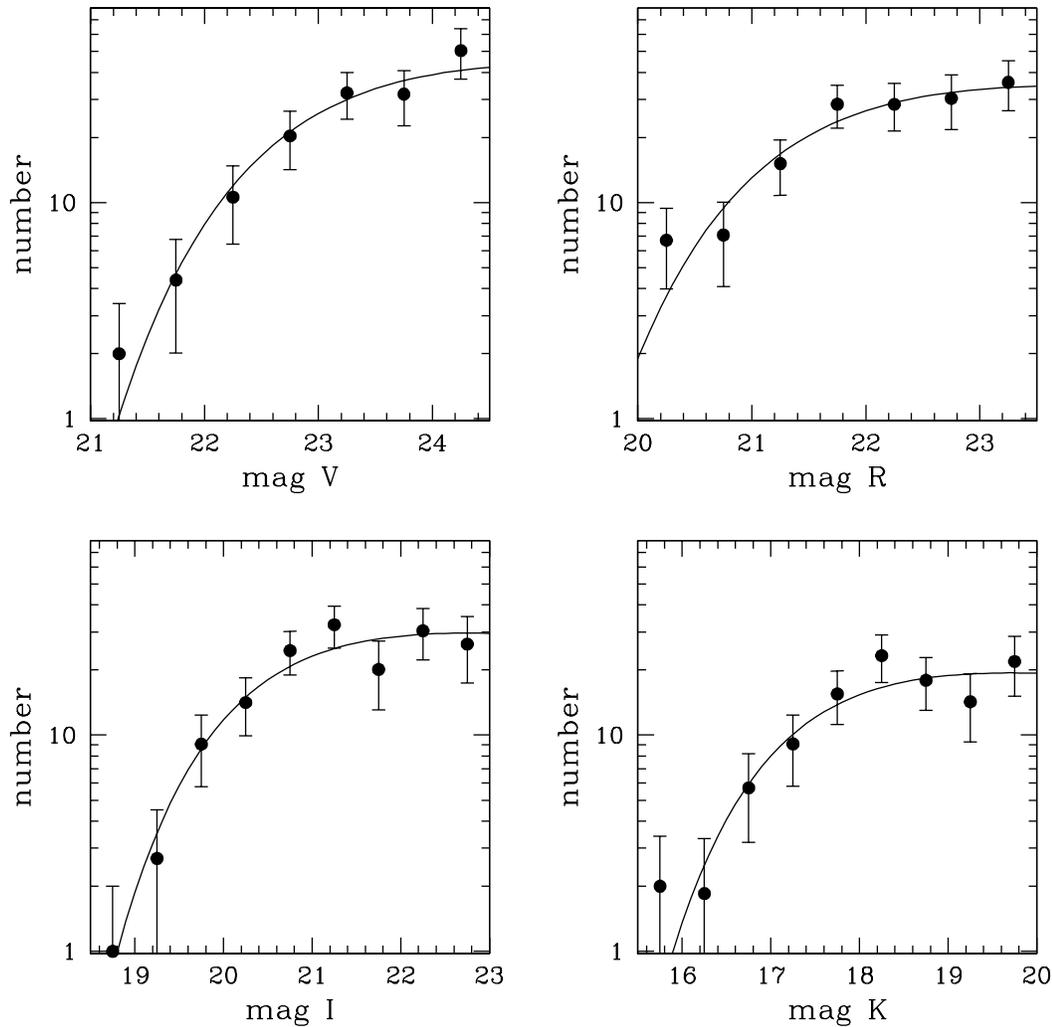}
\caption[]{ The CLFs of EIS\,0048 from V- to K-band (filled dots). In
each panel the best--fit Schechter function is also shown (solid
line).
\label{LUMFUN}
}
\end{figure}

To achieve a reasonable accuracy in the redshift estimate, we
considered only galaxies with signal-to-noise ratio $S/N > 5$ in at
least three bands. For each band, we computed the completeness
function of the sample as the ratio of the number of galaxies with
$S/N > 5$ in at least three bands to the total number of galaxies in
the catalog, as a function of the magnitude (see
Fig.~\ref{PSCOMP}). The CLFs have been computed in each band by
considering all the galaxies with photometric redshift up to the last
magnitude bin with completeness higher than $50\%$, as reported in
Table~\ref{TABCOMP}. Galaxy counts have been corrected for the
estimated incompleteness.

After the previous selection, field contamination is still present in
the peak of the redshift distribution around ${\rm z=0.64}$.  To
account for the remaining contamination, we took advantage of the
VIRMOS preparatory photometric survey as control sample (see
LBMI03). By selecting galaxies in the redshift range $z \in
[0.54,0.74]$ (LBMI03) in the VIRMOS control field, we obtained the
luminosity distribution of contaminant galaxies, and subtracted it
from galaxy counts in the same redshift range in the cluster field. At
$\mathrm{K} \leq 20$ the ratio of cluster to field galaxies without
any photometric redshift selection is $\sim 0.5$ (from Saracco et
al. \cite{SGC01} data in the CDF), whereas the same ratio after the
photometric redshift selection is $\sim 6.6$ (from the VIRMOS data),
thus proving the effectiveness of our procedure.

\subsection{Luminosity functions}\label{RES}

We modelled the CLFs with a weighted parametric fit of the Schechter
function, fixing the faint end slope to have $\alpha = -0.9$. We chose
not to fit $\alpha$ because the samples with photometric redshifts are
only complete to $M^{*} + 1.5\!--\!2.5$ (according to the
band). Moreover, $\alpha = -0.9$ is the value measured in the NIR for
the Coma cluster at $K < K^{*} + 3$ (De Propris et al. \cite{DES98})
and is also the value adopted by DPSE99 to study the behaviour of
$K^*$ as a function of the redshift. The errors on the CLFs were
computed by taking into account Poissonian fluctuations of cluster
galaxies counts and of the background counts in the redshift range $z
\in [0.54, 0.74]$. In Fig.~\ref{LUMFUN} we show the CLFs from V- to
K-band (filled circles) and the best--fitting Schechter functions
(solid lines).  The values of $M^{*}$ computed by the fits are listed
in Table~\ref{TABMSTAR}.

In the K-band we obtain $M^{*}=17.18 \pm 0.23$, in agreement within
the errors with the result of DPSE99 at the same redshift
($K^{*}=17.57 \pm 0.41$). By comparing the values of $M^{*}$ in I- and
K-band, we obtain $I^*-K^*= 3.06 \pm 0.28$ for the bright cluster galaxies
at $z=0.64$, in agreement with the findings of SED98 and NGZ01 at
similar redshifts.

\section{Luminosity evolution of bright ($M \leq M^*+1$) cluster galaxies}\label{DISC}

\begin{table}
\caption[]{The values of $M^{*}$ computed from the fits of the Schechter
function to the CLFs of EIS\,0048. $M^{*}$ is obtained by fixing $\alpha = -0.9$.}
\label{TABMSTAR}
$$
\begin{array}{cccc}
\hline
\noalign{\smallskip}
$$\mathrm{Band}$$ & $$\mathrm{M^*}$$\\
& $$\mathrm{mag}$$\\
\noalign{\smallskip}
\hline
\noalign{\smallskip}
$$\mathrm{V}$$ & 22.70 \pm 0.25\\
$$\mathrm{R}$$ & 21.27 \pm 0.20\\
$$\mathrm{I}$$ & 20.24 \pm 0.18\\
$$\mathrm{K}$$ & 17.18 \pm 0.23\\
\noalign{\smallskip}
\hline
\end{array}
$$
\end{table}

\begin{figure}
\includegraphics[angle=0,width=15cm,height=15cm]{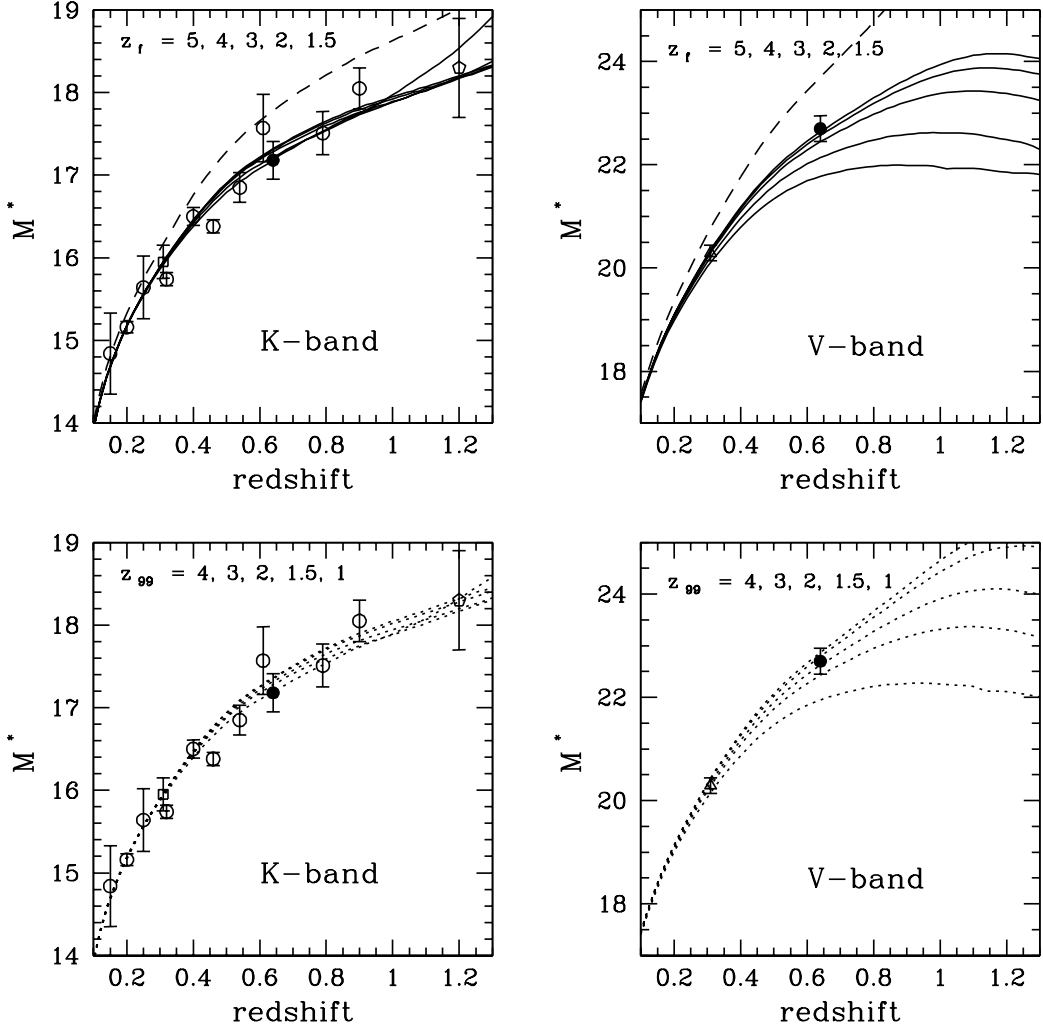}
\caption[]{ Evolution of the Schechter parameter $M^{*}$ in K- and
V-bands as a function of redshift. In the K-band the data come from
DPSE99 (empty circles), Andreon (\cite{AND01}, empty square), NKY01
(empty pentagon), and from this work (filled circle). In the V-band
the data come from Busarello et al. (\cite{BML02}, empty triangle) and
from this work (filled circle). In the upper panels, the dashed lines
represent the NE model.  The PLE models (solid lines) at different
formation redshift $z_f$ are shown in decreasing order of $z_f$ from
top to bottom.  In the lower panels the PLE models (dotted lines) are
shown as a function of the redshift $z_{99}$ for which where galaxies form 
the 99\% of their stars (decreasing from top to bottom).
\label{MSTAR}
}
\end{figure}

We discuss the luminosity evolution of bright galaxies in
clusters on the basis of the behaviour of the V- and K-band
characteristic magnitudes $M^*$ with redshift.  Complementary data for
the K-band are taken from DPSE99, Andreon \cite{AND01}, and NKY01, while for the
V-band are taken from Busarello et al. \cite{BML02}.  The different
data sets are represented in Fig.~\ref{MSTAR} with different symbols.
The points at $z=0.31$ in the K- and V-band are obtained from the
luminosity functions of Andreon (\cite{AND01}) and Busarello et
al. (\cite{BML02}), assuming $\alpha = -0.9$ up to $M \sim M^*+2$.  In
this way all the numerical estimates of $M^*$ have been obtained with
the same value of $\alpha$ and can be straightly compared.

In Fig.~\ref{MSTAR} no--evolution (NE, pure $k$--correction) and PLE
models are also shown. The galaxy models are constructed using Bruzual
\& Charlot (\cite{BrC93}) templates with a Scalo initial mass
function, solar metallicity, and an exponential star formation law
$e^{-t/\tau}$. We assumed the time scale $\tau =1$ Gyr, which is
suitable to reproduce the colour evolution of early type galaxies (see
Bruzual \& Charlot \cite{BrC93}; Merluzzi et al \cite{MBM03}).  The NE
predictions use model spectra with formation redshift $z_f = 5$ (the
redshift of the onset of SF). The predictions from PLE are given for
different $z_f$.  All the models are normalized to $K^*=10.90$ and
$V^*=14.17$ at the redshift of Coma (Bower et al. \cite{BLE92}; De
Propris et al. \cite{DES98}).

As widely discussed in literature (e.g. DPSE99; NKY01), the behaviour
of $K^*$ is inconsistent with a NE model. Our data points at $z=0.64$
confirm this result (at $\sim 2.5 \sigma$), and extend it in the
V-band (at $\sim 4 \sigma$).  The evolution in the K-band can be
described by PLE models at least up to $z \sim 1$ (the point at
$z=1.2$ in Fig.~\ref{MSTAR} should be taken with caution since NKY01
were not able to estimate the field contamination).  Since the K-band
samples the bulk of stellar mass, this result suggests that the
assembly of bright galaxies in clusters is complete at $z \sim 1$
(DPSE99).

The K-band data alone cannot discriminate among models with different
formation redshifts. At $z=0.64$ the V-band samples the UV restframe
and is therefore sensitive to even low levels of star formation
activity.  As can be seen in the upper right panel in
Fig.~\ref{MSTAR}, the evolution of $V^*$ is in agreement with PLE
models with $z_f \ge 3$.  This result depends on the time scale of
star--formation history, but we can use our point at $z=0.64$ in the
V-band to constrain the time when galaxies ceased to form stars. This
is possible because the magnitude evolution subsequent to this time is
almost independent from the preceding star--formation history.

In the lower panels of Fig.~\ref{MSTAR} we plot the PLE models
corresponding to the different epochs at which galaxies formed 99\% of
their present stars.  As it is apparent from the Figure, the
V-band data constrain the star formation in cluster bright galaxies to
end at $z \ge 1.5$.

\section{Summary and conclusions}\label{CONC}

We derived the galaxy LF in V-, R-, I-, and K-bands of the cluster
EIS\,0048 at $z \sim 0.64$ from new photometric observations carried out at 
ESO VLT with FORS2 and ISAAC. Cluster members have been selected with
the photometric redshift technique in order to enhance the contrast
among cluster- and background--foreground galaxies. The remaining
field contamination has been estimated as detailed in LBMI03.

The CLFs have been obtained up to $M-M^* \sim 1.5-2.5$ (according to
the waveband). We modelled the CLFs with a weighted parametric fit of
the Schechter function, fixing the faint end slope to be $\alpha =
-0.9$. The values of $M^*$ obtained in I- and K-bands are in agreement
with the analysis of DPSE99 and NGZ01 at similar redshifts.  We
collected results from literature and introduced PLE models with
different formation redshifts in order to discuss the evolution of
$M^*$ in the V- and K-bands.  The key factors driving the evolution of
bright galaxies in the scenario of hierarchical structure formation
are the epoch when the bulk of stellar populations is formed, the
cosmological time when mergers are effective to assemble the galaxies,
the amount of star formation induced by mergers, and the age of the
youngest stars.  Since the evolution of $K^*$ is consistent with a PLE
scenario at least up to $z \sim 1$, we can conclude that mergers must
have already assembled bright ($M \leq M^*+1$) cluster galaxies at
this redshift (see also DPSE99 for a thorough discussion).  At $z =
0.64$ the V-band samples the UV restframe wavelength region and is
sensitive to even low levels of SF.  We find that the SF in bright
cluster galaxies has to be almost completed at $z \sim 1.5$, whereas
the formation redshift is $z_f \geq 3$ assuming $\tau \sim 1$ Gyr as
the time scale for SF. These results lead to conclude that the
structure and the stars of bright cluster galaxies must have been
formed between $z=4\pm 1$ and $z=1.2\pm0.2$.

\begin{acknowledgements}

We are grateful to C. Lidman who helped us for the K-band photometric
calibration, and thank the ESO staff who effectively attended us
during the observation runs at Paranal.  We warmly thank the VIRMOS
Consortium who allowed us to use a subset of VIRMOS photometric data
base to estimate the field contribution in LBMI03. We thank the
unknown referee for his/her useful comments.  Michele Massarotti is
partly supported by a MIUR-COFIN grant.
\end{acknowledgements}

\end{document}